# Data-Driven Approach for Accelerating Selective Harmonic Elimination Algorithm in Parallel Power Converters

Ehsan Karimi, Shima Shahnooshi, Erfan Meshkati,Tomislav Dragičević, *Senior Member, IEEE,* and Frede Blaabjerg, *Fellow, IEEE*

*Abstract*—Current ripple minimization is one of the challenges in parallel converters to increase the capacitor lifetime in various applications. In this paper, a deep neural network-based phase-shifting (PS) technique is proposed for parallel-connected buck converters to minimize the amplitude of a selective harmonic component and facilitate a classic optimum PS at the same time. The proposed method identifies the global optimum point in real time, without the need for complicated computations. The common-link current, common-link voltage, and the duty ratios are selected as the inputs of the neural network to provide the proper phase shifts for the switching signals. To accumulate the required dataset, a Different Start-Same Step (DSSS) technique is also introduced to generate the training data and test/validation data in a separate way. The effect of the number of hidden layers on the network output error is investigated, and a proper number of hidden layers is designed based on a compromise between accuracy and computation efficiency (and execution time). Experimental results prove that the proposed artificial neural network-based PS method preserves the performance of classic optimum PS and minimizing the implementation time significantly.

*Index Terms*—Artificial Neural Network, Parallel Power Converters, Current Ripple Minimization, DC-DC power supply.

## I. INTRODUCTION

ARTIFICIAL Intelligence (AI) and particularly Artificial Neural Networks (ANN) have found numerous applications in various research areas and power electronics has not been excluded from this great breakthrough in data science [1]. Not only ANNs are able to extract information from raw data but also make it possible to parallelize and simplify the calculations. Hence, their applications in power electronics have significantly increased in recent years and presents numerous benefits including: i) turning the complex computations into simple and straightforward ones and modeling the nonlinear behavior of the controller. Therefore, it can lighten the calculation burden and reduce the controller power consumption [2]–[6]; ii) the ability to execute ANN with a parallel structure using field programmable gate arrays (FPGA) or application-specific integrated circuits (ASIC). Thus, calculations can be carried out agile and efficient. Regarding the above-mentioned merits, ANNs provide systems with the ability to either implement more calculations over a specific sampling period or increase the switching frequency [4], [7], [8]; iii) considering data acquired from real operating conditions when training ANN-based controllers may result in better system performance and robustness, compared with the mathematical modeling of the system. In other words, ANN-based controllers are capable of handling noise, non-idealities, and uncertainties; [9], [10] iv) improving the system transient response since ANNs are data-driven methods [10]; v) designing a common structure for different converters in such a way that merely the controller parameters should be modified based on the converter topology and parameters. It results in higher robustness, lower implementation time, and lower complexity [8], [11], [12]; vi) estimating system parameters and control objectives to improve the system performance and decrease the number of sensors [13]–[17]; vii) fault detection without detailed knowledge of the system model, and modify the control system under this condition [4], [18], [19], all in all enhancing the system efficiency, accuracy, and robustness.

To benefit from ANNs, previous studies have modeled either part of or the whole of conventional control systems with ANN-based controllers in power electronics applications in order to achieve one or more of the above-mentioned merits [6]. For instance, it has been utilized as a simplified substitute for reducing the computational burden in model predictive-based controllers. In [20], a multi-layer perceptron (MLP) classifier is trained by model predictive control (MPC) considering different load values and load types including linearities and nonlinearities for controlling full-bridge inverters. It leads to lower voltage total harmonic distortion (THD), independent relationship between load and controller parameters, and fast dynamic response. Likewise, a classification model has been used in [3], [5], [8] to emulate MPC for modular multilevel converters and other types of voltage source converters. Authors in [3] proposed an ANN-based estimator as a replacement for the exhaustive approach of determining the optimum weighting factors in finite control set-MPC. Moreover, the back-propagation ANN for DC-DC converters suggested in [7], [10] reduces the calculations and hence eliminates the limitations of using high sampling frequencies in MPC methods. To address the challenges of the conventional Proportional Integral (PI) controllers, such as parameter tuning, ANN-based

E. Karimi, S. Shahnooshi and E. Meshkati are with the Department of Electrical and Computer Engineering, Isfahan University of Technology, Iran (e-mail: (ehsan.karimi, s.shahnooshi, and e.meshkati)@alumni.iut.ac.ir).

T. Dragicevic is with the Center of Electric Power and Energy, Technical University of Denmark, Denmark (e-mail: tomdr@dtu.dk).

F. Blaabjerg is with the Department of Energy Technology, Aalborg University, Denmark (e-mail: fbl@et.aau.dk).

controllers have been investigated to eliminate cascade control loops in [21] and improve inverter waveforms for both steady-state and dynamic responses in [22]. ANNs are also exploited to enhance harmonic elimination in Distributed Static Synchronous Compensators (DSTATCOM) [9], [23]. Furthermore, ANNs have been among the control strategies in motor drive applications used as rotor parameters and speed estimator in [14]. It is been also utilized as a disturbance removal at low speeds to boost the performance of the stepper motor in [16]. Although some applications of ANN in power electronics have been addressed in the literature, the facilitation of optimization problems in power electronics is still an open research area.

Due to the superiorities of multiphase and parallel converters compared to a single phase converter in terms of overall cost and power density [24]–[26], various topologies of Multi-Input Single-Output (MISO), Single-Input Multi-Output (SIMO), and Multi-Input Multi-Output (MIMO) converters are being adopted for a wider range of applications including microprocessors and portable devices, photovoltaic systems, distributed power generation systems, charging circuits of H-bridge multilevel converters with non-symmetrical DC-link voltages, and chargers for electric vehicles [27]–[34]. Despite the advantages of parallel converters, the current of a common link is the sum of the phase currents with a higher ripple than the individual phases, resulting in higher EMI noise, lower reliability, and bulkier filter components [4], [24], [26], [35]. As a solution, ripple reduction can be accomplished based on phase-shifting (PS) among the phase currents. In symmetric parallel converters, ripple reduction is implemented using even PS. However, it is not applicable in practice because of different operating points of the parallel converters or having an asymmetric topology. Under this condition, solving an optimization problem is required [30], [36], [37].

Two approaches have been studied in the literature to perform optimum PS. In the first approach, phase shifts are calculated offline in the design stage and then applied to the converters without any modification in the operating stage, and hence these methods are not proper for applications with variable operating points [36]. Consequently, online optimization methods are required in such cases. In online methods, the optimization problem aiming at the optimum PS calculation is performed instantaneously as a function of the operating point in centralized and decentralized strategies [28], [36], [38], [39]. Centralized methods can perform well for interconnected power converters, such as the ones that are used in computer power systems, multi-string photovoltaic converters, and battery chargers [36], [37]. The main concerns of online methods are high computations in real-time operation and the time-consuming approach of obtaining the optimum solution. In iterative-based methods, such as Newton, gradient-based, and perturb and observe methods, the optimum PS for the new operating point is obtained after several consecutive switching ratios. Also, sometimes the optimization algorithm converges to a local optimum instead of the global one, which is not desirable. Although the proposed method in [37] guarantees the global optimum for almost all operating conditions, calculation time rises as the number of converters increases. It is due to the fact that the calculations are conducted in a serial structure.

As a result, it limits the switching frequency in applications with a high number of converters. It is deduced here that a good proposed method should calculate the global optimum PS in a short amount of time.

In this paper, an ANN-based regression model is exploited to reduce the complexity of optimum PS algorithm in the converter topologies with parallel/series connected inputs or outputs. This approach guarantees the global optimum PS for all steady-state operating points, while the latency time and the number of phases are completely independent. In addition to this, the calculations are facilitated and executed in parallel when compared to the conventional optimization methods. Also, a straightforward design and implementation are achieved. The performance of this method is verified in a SIMO topology constituted of buck converters, which is widely used in portable devices. The ANN-based phase shift controller is trained with the duty ratios, the load currents of different phases, and common-link voltage to estimate the optimum phase shifts of the parallel buck converters.

The remainder of this paper is organized into the following sections. In section II, the concept of conventional optimum PS is described for SIMO buck converters. According to the principles expressed in the preceding stage, the required dataset is generated to implement the proposed ANN-based PS using a proper network architecture which is descried in section III. Hardware implementation results are presented in section IV to verify the performance of the ANN-based PS method. Finally, conclusions are given in section V.

## II. CONVENTIONAL OPTIMUM PS METHOD

In general, in multi-phase converters with a common link, a classical controller is required to control the objective parameters of each phase separately. For instance, in portable devices, several converters are connected to a common DC-link providing power supply with various output voltage levels for different building blocks. One of the approaches adopted in the literature is duty ratio modification to fulfill the control objective. Fig. 1 shows a multi-phase structure consisting of three buck converters for the application of portable devices. The output voltages are controlled using PI controllers that produce proper duty ratio for each converter. To reduce the ripple of the common-link current, instead of direct use of the obtained duty ratios, they should be shifted, and therefore a PS calculator is required. The variable inputs of the PS calculator are the duty ratios determined by the operating condition, the average of load current given by the load power, and the common-link voltage (e.g. battery voltage). Also, the constant inputs, such as switching frequency, are selected during the design stage.

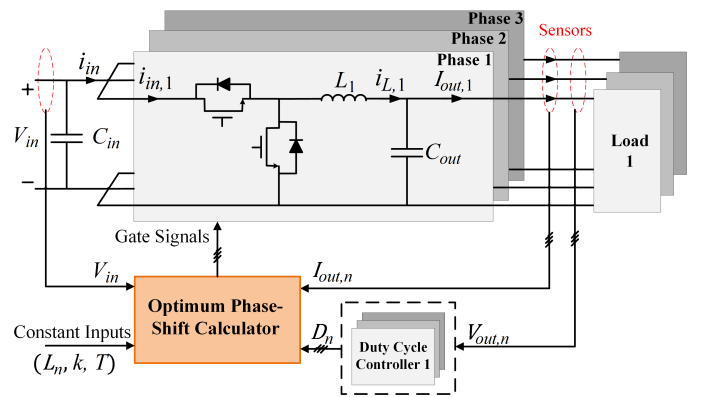

Fig. 1. General control structure of a multi-phase buck converter with optimum phase-shifting.

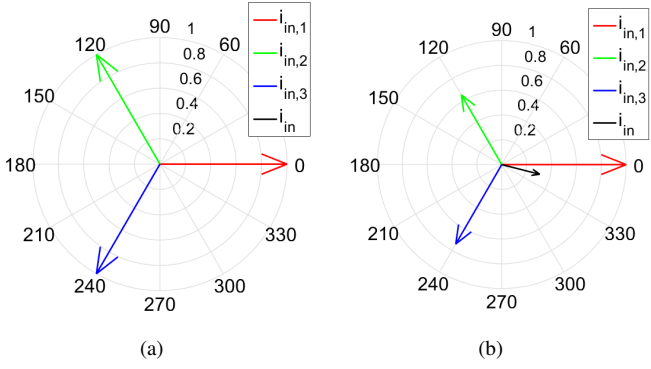

Fig. 2. Effect of the even phase-shifting (PS) technique on the summation vector of converters' current vectors in a system consisting of three paralleled-input buck converters under (a) symmetric operating condition and (b) asymmetric operating condition.

In symmetric multi-phase converters, the current ripple minimization is performed using even PS [37], in which the switching functions are shifted as follows:

$$\phi_{0,n} = (n-1)\frac{360}{N} \quad (1)$$

where $N$ is the total number of phases, $n$ denotes the phase number, and $\phi_{0,n}$ is the required phase shift for the converter existing in phase $n$.

Despite reducing current ripple in asymmetric parallel converters after implementing even PS, selecting more effective PS is necessary for minimizing the current ripple. Fig. 2 shows the ability of even PS to reduce the most effective harmonic component of common-link current ($i_{in}$). It is obvious that the summed vector is equal to zero under symmetric operating conditions, as shown in Fig. 2 (a). Whereas, under asymmetric conditions, the summed vector is not minimized using even PS as it can be seen in Fig. 2 (b). Therefore, the optimum PS technique should be adopted in order to reduce the input current distortion by minimizing the summation of the input current vectors of different converters.

Considering the buck converter in Fig. 1, the waveforms of input current ($i_{in,n}(t)$) and inductor current ($i_{L,n}(t)$) are shown in Fig. 3. It can be seen, the input current ripple is the same as inductor current ripple during the ON-state interval and it is calculated as:

$$\Delta i_{L,n} = \frac{(1-D_n).D_n.V_{in}}{L_n.f_{sw}} \quad (2)$$

where $\Delta i_{L,n}$, $D_n$, $f$, and $L_n$ are the inductor current ripple, duty ratio, switching frequency, and inductance of phase $n$, respectively, and $V_{in}$ is the common input of all phases. The

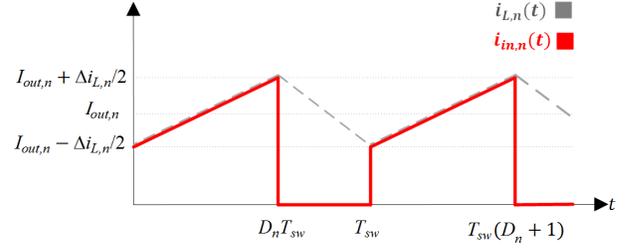

Fig. 3. Input ($i_{in,n}$) and inductor ($i_{L,n}$) currents of a buck converter.

time-domain equation of the input current over one switching period is represented as follows:

$$i_{in,n}(t) = \begin{cases} \frac{\Delta i_{L,n}}{D_n T_{sw}}.t + I_{out,n} - \frac{\Delta i_{L,n}}{2}, & 0 \leq t \leq D_n T_{sw} \\ 0, & D_n T_{sw} < t < T_{sw} \end{cases} \quad (3)$$

where $I_{out,n}$ is either the average value of inductor current or DC output current of the converter in phase $n$ and $T_{sw}$ is the switching period.

In a parallel structure of three buck converters, the input currents of the phases are summed up producing the resultant input current as:

$$i_{in}(t) = \sum_{n=1}^{N} i_{in,n}(t) \quad (4)$$

where $i_{in,n}(t)$, is the input currents of phase number $n$, and $i_{in}(t)$ is the resultant input current. It means that the input current ripple of individual phases yields the magnitude of the resultant current ripple that can be reduced by conducting harmonic component minimization based on optimum phase-shifting. To do so, the harmonic components of the input current of each phase should be described in the frequency domain by obtaining the Fourier series as follows:

$$i_{in,n}(t) = \frac{a_{n0}}{2} + \sum_k [a_{nk}\cos(k(\omega t - \phi_{0,n})) + b_{nk}\sin(k(\omega t - \phi_{0,n}))] \quad (5)$$

where $k$ is the harmonic order, and $\omega$ is the angular frequency. The parameter $a_{n0}$ denotes the DC component, and the parameters $a_{nk}$ and $b_{nk}$ are the coefficients of sinusoidal functions and they are obtained as:

$$a_{n0} = I_{out,n}.D_n \quad (6a)$$

$$a_{nk} = \frac{1}{k\pi}((I_{out,n} + \frac{\Delta i_{L,n}}{2}).\sin(k\omega D_n T_{sw}) + \ldots \\ \frac{\Delta i_{L,n}}{k\omega D_n T_{sw}}.[\cos(k\omega D_n T_{sw}) - 1]) \quad (6b)$$

$$b_{nk} = \frac{1}{k\pi}(\frac{\Delta i_{L,n}}{k\omega D_n T_{sw}}.\cos(k\omega D_n T_{sw}) - \ldots \\ (I_{out,n} + \frac{\Delta i_{L,n}}{2}).\cos(k\omega D_n T_{sw}) + (I_{out,n} - \frac{\Delta i_{L,n}}{2})) \quad (6c)$$

To represent the Fourier series in a simpler form, the sinusoidal functions of (3) are merged to obtain an alternative that is written as:

$$i_{in,n}(t) = \frac{a_{n0}}{2} + \sum_k [A_{nk}\cos(k(\omega t - \phi_{0,n})) - \phi_{nk}] \quad (7)$$

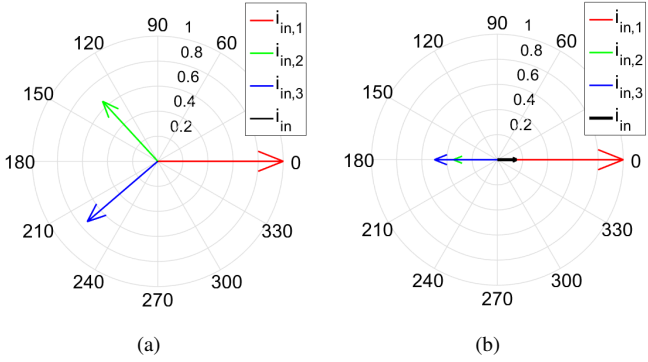

Fig. 4. Selective harmonic minimization using optimum PS, for (a) valid and (b) invalid inputs of inverse cosine functions.

where $A_{nk}$ and $\phi_{nk}$ are defined using the coefficients $a_{nk}$ and $b_{nk}$ given as:

$$A_{nk} = \sqrt{a_{nk}^2 + b_{nk}^2} \tag{8}$$

$$\phi_{nk} = \arctan(b_{nk}, a_{nk}) \tag{9}$$

where the returned $\phi_{nk}$ is between $-\pi$ and $\pi$. Using phasor transformation and the summed input current in (2), the resultant input current can be expressed as a phasor:

$$A_k e^{-j\theta_k} = \sum_{n=1}^{N} A_{nk} e^{-j\theta_{nk}} \tag{10}$$

where $\theta_{nk}$ is calculated as:

$$\theta_{nk} = k\phi_{0,n} + \phi_{nk} \tag{11}$$

To ease the expression of the optimum PS, three buck converters with parallel inputs connected to a common DC-link are considered, i.e. $N = 3$. However, the process can be generalized to other types of multiphase structures with any number of converters. The main purpose of the current ripple reduction in asymmetric multiphase converters is to minimize the summed phasor in (10) for a specific harmonic component. Note that it is possible to minimize every selected harmonic component by the proper selection of $k$. If it is assumed that the value of $A_k$ equals zero, and (10) can be written based on sinusoidal functions, the following equations are obtained. The phase of the current vector of the first converter, i.e. $\theta_{1k}$, is selected zero to be the reference of the two other phase shifts. By substituting (11) into (10), and then minimizing the summed current phasor, the required phase shift of the switching functions is obtained as:

$$\phi_{0,1} = \frac{-\phi_{1k}}{k} \tag{12a}$$

$$\phi_{0,2} = \frac{1}{k}[\arccos(\frac{1}{2} \cdot \frac{A_{3k}^2 - A_{2k}^2 - A_{1k}^2}{A_{1k}.A_{2k}}) - \phi_{2k}] \tag{12b}$$

$$\phi_{0,3} = \frac{1}{k}[2\pi - \arccos(\frac{1}{2} \cdot \frac{A_{2k}^2 - A_{1k}^2 - A_{3k}^2}{A_{1k}.A_{3k}}) - \phi_{3k}] \tag{12c}$$

Notably, the derived approach of calculating phase shifts in (11) is valid only when the inputs of inverse cosine functions are in the interval [-1,1]. In this case, the entire cancellation of the corresponding harmonic component is possible, as shown in Fig. 4(a). Otherwise, the magnitude of $k^{th}$ harmonic component can be minimized to a non-zero value by adopting $0°$ and $180°$ phase shifts to the angle of current phasors ($\theta_{nk}$). First, the current phasor with the highest magnitude is selected among the three feasible current phasors, and then its switching function is calculated by setting the angle of the corresponding current phasor to $0°$ in (11). Whereas, the switching functions of two other phases are calculated using a $180°$ phase shift to perform the minimization of the associated harmonic component. Fig. 4(b) illustrates the optimum PS when the inputs of inverse cosine functions in (12) are invalid.

The whole process of optimum PS for the structure depicted in Fig. 1 is expressed as a pseudo code in Algorithm 1. The variable inputs $D_n$, $I_{out}$, and $V_{in}$ depend on the operating condition, while $D_n$ is determined by the duty-ratio controller, and $I_{out}$ is dictated by the load power. The constant inputs are $k$, $L_n$, and $T$ that are selected during the design stage. Due to the significant effect on the current ripple and filtering requirements, minimizing the first harmonic order, i.e. $k = 1$, is usually the best choice. In the next step, the Fourier series coefficient for the selected harmonic order can be calculated. Then, according to the inputs of inverse cosine functions, the optimum phase-shifts are computed and applied to the gate pulses of the semiconductor devices.

## III. PROPOSED ANN-BASED PS METHOD

ANN-based system modeling is exploited as a technique to model either a defined system (white-box) or an undefined system (black-box) using the system outputs corresponding to a specific set of input data [40]. In this section, a supervised learning method is investigated for the purpose of ANN-based modeling of the optimum PS calculator for the reduction of common-link current ripple in parallel converters. First, the required dataset is generated based on the concept of DSSS technique in the first stage. Then, the weights of the proposed ANN are tuned using the backpropagation learning algorithm to minimize the cost function consisting of ANN outputs and

---

**Algorithm 1** Pseudo Code of the Optimum PS calculator for three paralleled-input buck converters

**Input:** $L_n$, $T_{sw}$, and $k$ from design stage, $D_n$ from output voltage controller, $I_{out,n}$ from load power, $V_{in}$ from battery.
    **for** $n =1$ to $3$ **do**
        Calculate $\Delta i_{L,n}$ using Eq.2
        Calculate $a_{nk}$ and $b_{nk}$ using Eq.6
        Calculate $A_{nk}$, $\phi_{nk}$ using Eq.8 and Eq.9
    **end for**
    **if** $(arg(cos^{-1}(...)) \leq 1)$ **then**
        Calculate $\phi_{0,1}$, $\phi_{0,2}$, $\phi_{0,3}$ using Eq.12
    **else**
        $n_{max} \leftarrow argmax(A_{1k}, A_{2k}, A_{3k})$
        $\theta_{n_{max}k} \leftarrow 0°$
        $\theta_{nk} \leftarrow 180°, n \neq n_{max}$
        Calculate $\phi_{0,1}$, $\phi_{0,2}$, $\phi_{0,3}$ using Eq.11
    **end if**
**Output:** $\phi_{0,1}$, $\phi_{0,2}$, $\phi_{0,3}$

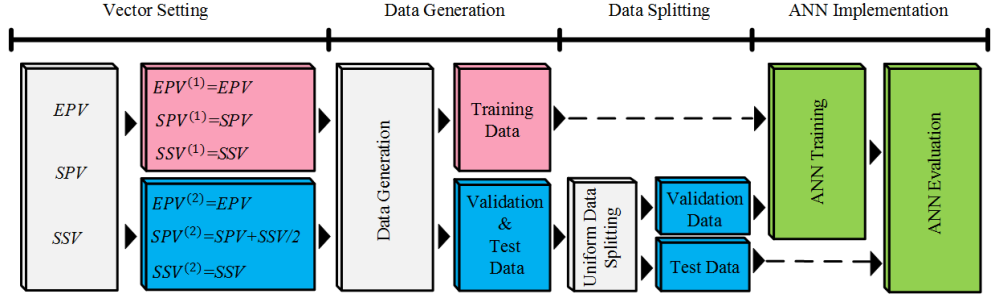

Fig. 5. The pipeline of data generation using introduced DSSS technique for ANN-based controller implementation. EPV: Ending Point Vector, SPV: Starting Point Vector, and SSV: Step Size Vector.

targets during the training process. Finally, the trained ANN is implemented on a desired hardware in real operating condition. The required modeling stages are expressed in the following.

### A. Dataset Generation

The acquisition and storage of system outputs for a specific input dataset are conducted in this stage. Note that the model inputs can be either similar to or different from system inputs that are stored alongside their associated system outputs to be used in the following stages.

Two methods have commonly been employed in previous works as presented in section II for the purpose of dataset acquisition, which are briefly explained in the following:

- **Data logging using experimental implementation:** In this method, the system parameters are measured, and the dataset required to train ANNs are logged while the system is operating in real condition.
- **Data generation using simulated system:** To generate data in this method, the corresponding system is implemented in a simulation environment, such as Simulink/MATLAB, and then the ANN-training Dataset are obtained.

In both methods, the system operates under different operating conditions, and the associated parameters are generated and stored to train the ANN-controller in the next steps. However, these methods lead to high processing time, and it is a time-consuming procedure for the purpose of generating the required dataset.

In this paper, a mathematical approach is adopted to facilitate the process of data acquisition and reduce the required computation using the method in the presented diagram in Fig. 5. The concept of DSSS is utilized in this stage to generate two distinct series of data. After selecting a proper step size and two different starting points, Algorithm 1 is executed two times to generate separate datasets. The first dataset is used to train the ANN, while the second one is used throughout the validation and test process. The principles of the DSSS technique is described in the following.

As depicted in Fig. 5, the first step is determining the maximum and minimum values of the inputs of the optimum PS calculator. In this regard, the Starting Point Vector ($SPV$) contains the minimum values of the output currents and duty ratios, while the Ending Point Vector ($EPV$) includes the maximum ones, which are defined as:

$$SPV = \left[I_{out,1}^{min}, I_{out,2}^{min}, I_{out,3}^{min}, D_1^{min}, D_2^{min}, D_3^{min}, V_{in}^{min}\right] \quad (13a)$$

$$EPV = \left[I_{out,1}^{max}, I_{out,2}^{max}, I_{out,3}^{max}, D_1^{max}, D_2^{max}, D_3^{max}, V_{in}^{max}\right] \quad (13b)$$

Considering the application of three paralleled-input buck converters, for example in portable devices, 9 and 12.6 are selected as the minimum and maximum values for $V_{in}$. Also, the output currents are within 0.2 A and $1.1 I_{nom}$ A as the lowest and highest values, where $I_{nom}$ is the nominal output current. In steady-state operating condition, the value of $D_n$ is equal to the nominal value during the whole process. The next stage is to determine a proper Step Size Vector ($SSV$), that is set to 0.1 and 0.05 A for common-link voltage and output currents, respectively. Note that, the smaller step size increases data volume and thereby the precision of ANN controller. However, it can have an undesired effect on training time. In general, the length of dataset ($N_{data}$) can be calculated as:

$$N_{data} = \prod_{m=1}^{7} \frac{EPV[m] - SPV[m]}{SSV[m]} \quad (14)$$

where $m$ refers to the location number of vectors elements.

As shown in Fig. 5, two distinct datasets are generated in two separate executions. In the first dataset used to train ANN, the values of $SPV^{(1)}$, $EPV^{(1)}$, and $SSV^{(1)}$ are equal to $SPV$, $EPV$, and $SSV$, respectively. The second dataset is generated for testing and validation purposes, and the corresponding vectors are defined $SPV^{(2)}$, $EPV^{(2)}$, and $SSV^{(2)}$. The two latter vectors have similar values compared to training vectors, while $SPV^{(2)}$ is set to $SPV + \frac{SSV}{2}$. As a result, the training data and the test/validation data are entirely independent.

Next, the second dataset is uniformly split to produce test and validation data. Training and validation data are exploited in the process of ANN training, and finally, test data are used to evaluate the performance of trained ANN.

### B. Proposed ANN Architecture

MLP or Fully-Connected Neural Network (FCNN), as a Feedforward Neural Network (FNN), has become one of the attractive alternatives for controller modeling in power electronics applications [3], [5], [7], [8], [10], [20]–[22]. It is mainly due to its fast processing time in obtaining the outputs, when compared to Convolutional Neural Network (CNN). Fig. 6 shows a general structure of FCNN that consists of input layer, hidden layers, and output layer. Neurons are the constituents of each layer and have several inputs and outputs. The inputs of each neuron are the outputs of all neurons existing in the previous layer, and the output of the corresponding neuron is obtained as:

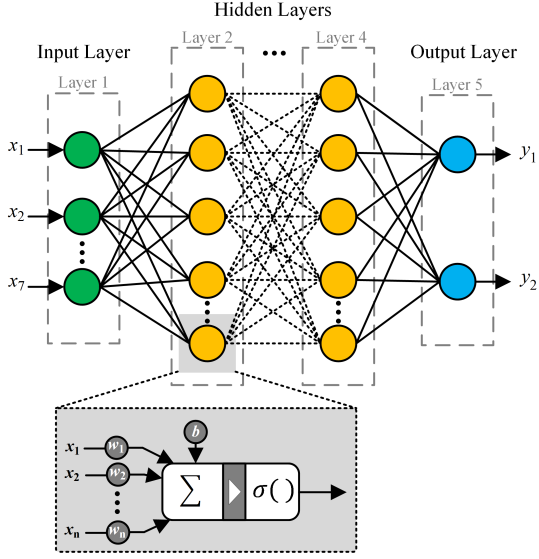

Fig. 6. Proposed FCNN architecture for calculating phase shifts in three parallel buck converters.

TABLE I
DETAILS OF THE PROPOSED ANN MODEL ARCHITECTURE (SEE FIG. 6)

| Layer number | Layer type | Operation type | Hyper parameter |
| --- | --- | --- | --- |
| 1 | Input layer | Normalization | 7 (neurons) |
| 2 | Hidden layer | Dense ReLU function Dropout | 20 (neurons) — 5% (drop ratio) |
| 3 | Hidden layer | Dense ReLU function Dropout | 20 (neurons) — 5% (drop ratio) |
| 4 | Hidden layer | Dense ReLU function Dropout | 20 (neurons) — 5% (drop ratio) |
| 5 | Output layer | Dense ReLU function | 2 (neurons) — |

$$A = \sigma(W^T X + b) \quad (15)$$

where $X$, $W$, $b$, and $\sigma$ are the input matrix of the neuron, the weight matrix, the bias value, and the activation function, respectively. $W$ and $b$ are trainable parameters that are trained during the training stage. Also, to take advantage from the neurons ability to model nonlinear systems, a nonlinear activation function should be adopted.

The architecture of the established ANN is shown in Table I. The input data with different orders of magnitude are scaled to [0, 1] in the input layer to improve the network performance. Next, the scaled data enter the first hidden layer with a Dense structure. Due to the fact that the proposed ANN should be implemented over a short time on a controller IC with limited resources, a simple activation function is required. Therefore, Rectified Linear Unit (ReLU) activation function is used for all layers and defined as:

$$ReLU(x) = max\{0, x\} = \begin{cases} x, & x \geq 0 \\ 0, & x < 0 \end{cases} \quad (16)$$

This activation function presents benefits in terms of high implementation speed, as well as their derivatives, and they can speed up the algorithm convergence. Also, a Dropout technique with the ratio equalling 0.05 is employed after each hidden layer to avoid overfitting.

Two neurons are located in the output layer that provide the values of $\phi_{0,2} - \phi_{0,1}$ and $\phi_{0,3} - \phi_{0,1}$ over the range of [0,1]. Considering $\phi_{0,1} = 0$, the ANN outputs are allocated to $\phi_{0,2}$ and $\phi_{0,3}$, and then scaled to a degree range.

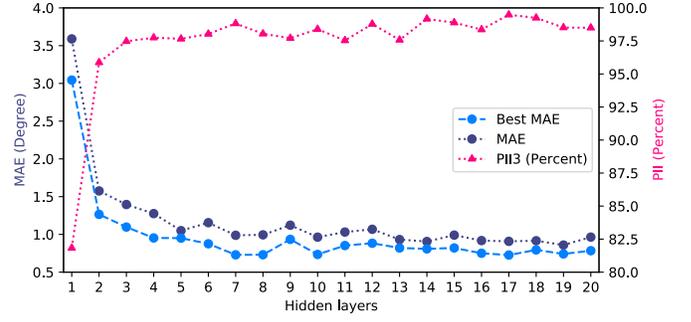

Fig. 7. ANN model output error for different number of hidden layers. PII represents the model's accuracy after quantizing its outputs using a PWM generator with 3-degree accuracy.

### C. ANN Training

At this stage, the ANN with the proposed architecture, as described in Table I, is trained by the datasets generated by the DSSS technique. ANN training is implemented in the Keras Deep Learning Application Programming Interface (API). Cost function is Mean-Squared Error (MSE) that defined as:

$$MSE = \frac{1}{n}\left(\sum_{i=1}^{n}(Y_i - \widehat{Y_i})^2\right) \quad (17)$$

where $n$, $Y$, and $\widehat{Y}$ denote the data number, the predicted output values, and the target values, respectively.

The proposed ANN is trained over 10 consecutive, independent procedures. The average value of MSE is obtained $1.73 \times 10^{-4}$ and the value of Mean Absolute Error (MAE) is reported $1.09°$ for $\phi_{0,2}$ and $\phi_{0,3}$. Also, the absolute value of the ANN output error is within $3°$ for more than $97.4\%$ of the existing test data.

In a similar way, the values of MAE and best MAE are shown in Fig. 7 by changing the number of hidden layers from 1 to 20. In this study practical condition, the digital counter speed is considered 120 times higher than the PWM frequency when implementing PWM generator. Therefore, the accuracy is $3°$, and $PII3$ refers to the percentage of test data with the output error less than $3°$. As it can be seen from Fig. 7, $PII3$ experiences an increase as the number of hidden layers rises. Furthermore, an ANN structure with higher number of hidden layers results in lower values of MAE and the best MAE, and therefore better performance of the corresponding ANN. However, the trade-off between ANN output error and hardware requirements determines the optimum number of hidden layers. To clarify, the ANN training results are reported in Table II up to six hidden layers. Notably, the input and first hidden layers are merged in this work to reduce computations, and the weights and biases of the new layer are calculated as:

$$w_{q,i}^{new} = \frac{w_{q,i}^{(2)}}{(X_{i,max} - X_{i,min})} \quad (18a)$$

TABLE II
ANN TRAINING RESULTS FOR DIFFERENT NUMBER OF HIDDEN LAYERS

| Hidden layers | MSE | MAE (Degree) mean | MAE (Degree) best | FLOPs | Parameters | Horizon implementation time (FPGA clock) | Percent In Interval 3° (PII3) |
|---|---|---|---|---|---|---|---|
| 1 | $6.24 \times 10^{-4}$ | 3.59 | 3.04 | 969 | 202 | 2 | 82.5 |
| 2 | $2.32 \times 10^{-4}$ | 1.57 | 1.26 | 1124 | 622 | 3 | 96.0 |
| 3 | $2.19 \times 10^{-4}$ | 1.39 | 1.09 | 3049 | 1042 | 4 | 97.4 |
| 4 | $1.73 \times 10^{-4}$ | 1.27 | 0.95 | 5775 | 1462 | 5 | 97.6 |
| 5 | $1.42 \times 10^{-4}$ | 1.05 | 0.95 | 9302 | 1882 | 6 | 97.5 |
| 6 | $1.97 \times 10^{-4}$ | 1.15 | 0.87 | 13630 | 2302 | 7 | 97.9 |

TABLE III
SYSTEM PARAMETERS USED FOR EXPERIMENTAL VERIFICATION

| Parameter | Value n=1 | Value n=2 | Value n=3 |
|---|---|---|---|
| $V_{in}$: Input voltage (V) | 12.6 | 12.6 | 12.6 |
| $V_{out,n}$: Rated output voltage (V) | 5 | 2.5 | 3.3 |
| $P_{out,n}$: Rated output power (W) | 10 | 5 | 5 |
| $C_{out,n}$: Output Capacitor ($\mu F$) | 270 | 1330 | 420 |
| $L_n$: Inductance ($\mu H$) | 63.4 | 48.1 | 76.3 |
| $f_{sw}$: Switching frequency (kHz) | 200 | 200 | 200 |

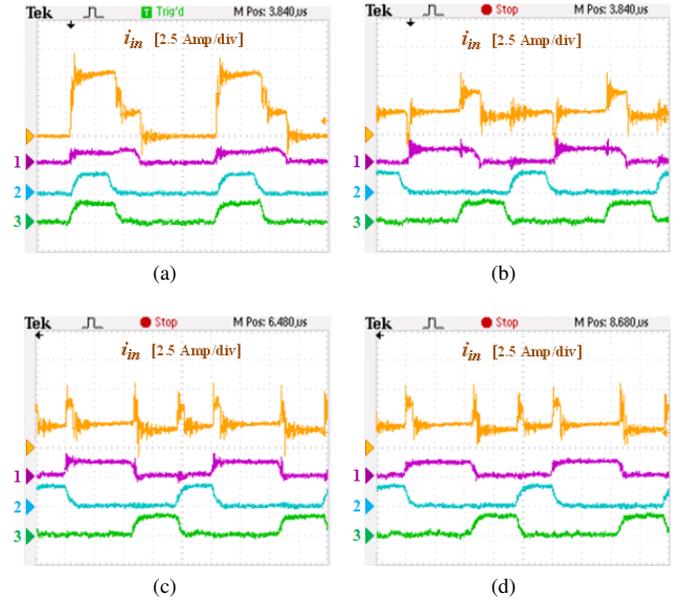

$$b_q^{new} = b_q^{(2)} - \sum_{i=1}^{7} \frac{w_{q,i}^{(2)} * X_{i,min}}{(X_{i,max} - X_{i,min})} \quad (18b)$$

where $w_{q,i}^{(2)}$ and $w_{q,i}^{new}$ are the weight values corresponding to the $i^{th}$ input of $q^{th}$ neurons in the first hidden layer and new layer, respectively. Also, $b_q^{new}$ and $b_q^{(2)}$ are the biases of the $q^{th}$ neurons in new and first hidden layers, respectively. The parameters $X_{i,max}$ and $X_{i,min}$ refer to the maximum and minimum values of the $i^{th}$ input feature in the dataset.

While reducing MSE and MAE parameters by adding more hidden layers, floating point operations (FLOPs), network parameters, and FPGA clock increase linearly. It means that higher number of hidden layers leads to better network accuracy but higher calculation burden and storage requirements and therefore increases the FLOPs. As stated in [7], if FPGA is used for implementing the corresponding algorithm, the implementation time increases by one clock when adding one hidden layer. Therefore, there is a compromise between the network accuracy and the existing hardware capacity when selecting the number of hidden layers.

## IV. HARDWARE IMPLEMENTATION

To verify the performance of the ANN-based PS calculator, the proposed ANN is implemented on the prototype of three paralleled-input buck converters using FPGA. The selected parameters of the experimental system are selected considering the application of portable devices, such as laptops, and are reported in Table III. Also, IRF1404 MOSFET is used as the switching devices in the structure of buck converters.

Fig. 8 shows the switching triggering signals of three parallel converters and the common-link current under full-load operating condition without PS, and also in the even PS, optimum PS, and ANN-based PS methods. The values of duty ratios are actively controlled by PI controllers to set the output voltages to their rated values (see Table III). The common-link currents after applying the four mentioned methods are also depicted in Fig. 8. As it can be seen from the comparison of Fig. 8(c) and Fig. 8(d), The proposed method presents the same results as the calculation-based PS with good precision. It is apparent that PS values do not effect the duty ratios, as they are independently controlled by the main control loop.

Fig. 8. Experimental results of the switching signals of buck converters 1 to 3 (with indicators 1 to 3, respectively) and the common-link current (orange waveform) under full-load operating condition in (a) no PS, (b) even PS, (c) optimum PS, and (d) ANN-based phase shifting. Time scales are 1 $\mu s/div$.

TABLE IV
COMPARISON AMONG DIFFERENT PS METHODS

| Methods of PS | $A_{in1}$ (Amp) Full-load | $A_{in1}$ (Amp) Half-load | FPGA clock |
|---|---|---|---|
| No PS | 2.93 | 1.45 | 0 |
| Even PS | 0.67 | 0.34 | 0 |
| Optimum PS | <0.01 | <0.01 | >13 |
| **Proposed ANN-based PS** | **<0.01** | **<0.01** | **4** |

The harmonic spectrum of the common-link current corresponding to the four above-mentioned PS methods is shown in Fig. 9. It contains the amplitudes of low-order harmonics, and it is apparent that the amplitude of the first harmonic order is obtained around zero in both optimum PS and proposed ANN-based PS methods.

The comparison results of the first harmonic amplitude

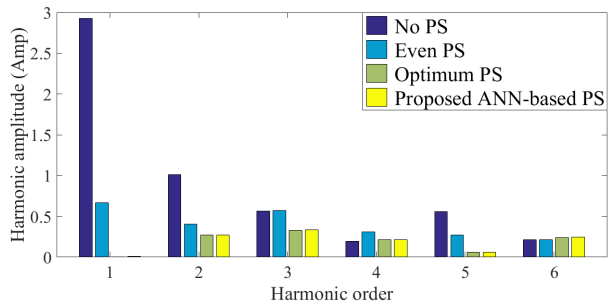

Fig. 9. Different PS methods effectiveness on low-order harmonic amplitudes.

($A_{in,1}$) and FPGA clock required to execute no PS, even PS, optimum PS, and proposed ANN-based PS methods are shown in Table IV. As it can be seen, the effect of proposed ANN-based method on $A_{in,1}$ is relatively the same as the optimum PS. However, by employing ANN, the implementation time has been reduced from more than 13 FPGA clocks to 4 FPGA clocks. The rms value of the common-link current is also reported 3.21 Amp, 2.18 Amp, 2.1 Amp, and 2.1 Amp for no PS, the even PS, the optimum PS, and the ANN-based PS methods respectively.

## V. Conclusion

This paper has proposed an ANN-based PS method to minimize the selective harmonic of the common-link current, thus reducing the ripple of the input current. The salient feature of the proposed ANN-based technique is that it enables the parallel and low-delay computations, thus facilitating the implementation on FPGA with a higher frequency and reducing the execution time considerably. Moreover, a mathematical approach is adopted to generate dataset for ANN-controller training and testing, resulting in low processing time. The proposed method was applied to three parallel-connected buck converters. After assessment of the amplitude of the first harmonic component through the experimental results, it is proved that the proposed method performs as well as the conventional optimum PS method in terms of harmonic cancellation. Also, the rms value of the input current is reduced when compared to the no PS approach. Notably, the proposed ANN-based control method can be extended to other types of multi-phase converters with any number of phases, which will be the focus in future work.